\documentclass[prd,twocolumn,10pt,aps,amsmath,amssymb,nofootinbib,superscriptaddress,showkeys,showpacs]{revtex4-1}
\usepackage{epsfig} 
\usepackage{amsmath} 
\usepackage{graphicx}
\usepackage{color}
\usepackage{float}
\usepackage{soul,xcolor}

\newcommand{\be}{\begin{equation}}
\newcommand{\ee}{\end{equation}}
\newcommand{\ba}{\begin{array}}
\newcommand{\ea}{\end{array}}
\newcommand{\bea}{\begin{eqnarray}}
\newcommand{\eea}{\end{eqnarray}}

\def\ket#1{\left| #1\right\rangle}

\begin{document}

\title{Quantum correlations in terms of neutrino oscillation probabilities}

\author{Ashutosh Kumar Alok}
\email{akalok@iitj.ac.in}
\affiliation{Indian Institute of Technology Jodhpur, Jodhpur 342011, India}

\author{Subhashish Banerjee}
\email{subhashish@iitj.ac.in}
\affiliation{Indian Institute of Technology Jodhpur, Jodhpur 342011, India}

\author{S. Uma Sankar}
\email{uma@phy.iitb.ac.in}
\affiliation{Indian Institute of Technology Bombay, Mumbai 400076, India}

\date{\today} 

\begin{abstract}
Neutrino oscillations provide evidence for the mode entanglement 
of neutrino mass eigenstates in a given flavour eigenstate. Given 
this mode entanglement, it is pertinent to consider the relation 
between the oscillation probabilities and other quantum correlations. 
In this work, we show that all the well-known quantum correlations, 
such as the Bell's inequality, are directly related to the neutrino 
oscillation probabilities.  The results of the neutrino oscillation
experiments, which measure the neutrino survival probability to be
less than unity, imply Bell's inequality violation.
\end{abstract}

\maketitle

%%%%%%%%%%%%%%%%%%%%%%%%%%%%%%%%%%%%%%%
\section{Introduction}
%%%%%%%%%%%%%%%%%%%%%%%%%%%%%%%%%%%%%%%
The foundations of quantum mechanics are usually studied in 
optical or electronic systems. In such systems, the interplay 
between the various measures of quantum correlations is well known. 
Inspired by the recent technical advances in high energy physics 
experiments, in particular the meson factories and the long 
baseline neutrino experiments, attention has also been directed 
towards subatomic physics \cite{meson,expt,neutri1,neutri2,neutri3,
neutri4,sbalok,Banerjee:2015mha,Blasone:2015lya,Formaggio:2016cuh}.

The study of quantum correlations in neutrinos, have been mostly 
focussed on entanglement \cite{neutri1,neutri2,neutri3,neutri4}. 
Recently, a temporal analogue of Bell's inequality,  
the Leggett-Garg inequality, has been studied in the context of neutrino oscillations \cite{Formaggio:2016cuh}. 
Here, along with entanglement we also study the other quantum 
correlations such as Bell's inequality violation, teleportation fidelity 
and geometric discord in the context of two flavor
neutrino oscillations as well as study the interplay between them. 
In particular, we show that all these quantities are directly 
related to the neutrino oscillation probabilities.

Neutrino oscillations are experimentally well established 
\cite{Bahcall:2004ut,Eguchi:2002dm,Araki:2004mb,Ashie:2004mr,Michael:2006rx,
Abe:2013hdq,Abe:2013fuq}.
Such oscillations are possible if both of the following conditions 
are satisfied:
\begin{itemize}
\item
The neutrino flavour state is a linear superposition of non-degenerate 
mass eigenstates.
\item
The time evolution of a flavour state is a coherent superposition of the
time evolution of the corresponding mass eigenstates.
\end{itemize}
The coherent time evolution implies that there is {\it mode 
entanglement} between the  mass eigenstates which make up a flavour 
state. Such mode entangled states have been the subject of intense discussions over the last two decades \cite{enk,gerry,vedral,exp1,exp2,barnum}, resulting 
in the general consensus of subspace entanglement as a generalized feature of inter particle entanglement \cite{barnum}.
It has been the subject of many theoretical and experimental proposals \cite{exp1} as well as successful experimental 
realizations \cite{exp2} in atom-photon systems. Here we use the concept of mode entanglement to relate 
flavour oscillations to bipartite entanglement
of single particle states.

The quest for understanding quantum correlations could be thought 
to have begun with the efforts of Einstein-Podolsky-Rosen (EPR) 
\cite{epr}. A quantitative understanding of EPR led to the development 
of Bell's inequality \cite{bell}, with refinements leading to the 
Bell-CHSH (Clauser-Horn-Shimony-Holt) inequalities \cite{chsh}. 
Violation of Bell's inequality quantifies the non-locality inherent 
in the system. A weaker, though very popular and widely studied facet 
of quantum correlations, is entanglement \cite{wootters}. This has been 
applied to understand the process of teleportation \cite{bennett}.  
A still weaker measure is quantum discord \cite{ollivier,henderson} 
and was developed as the difference between the quantum generalizations 
of two classically equivalent formulations of mutual information. 
Since states with are separable and hence have no entanglement could 
still have non zero discord, our present understanding of quantum 
correlations is that it is a complex entity with many facets. There 
is now an abundance of measures of quantum correlations such as 
quantum work deficit \cite{horo}, measurement induced disturbance 
\cite{luo1} and dissonance \cite{modi}.

In this paper we study a number of quantum correlations 
in the context of two-flavour neutrino oscillations. Among them are 
mode non-locality, concurrence, discord and teleportation fidelity. We 
find that all these quantum correlations are simple functions of 
the neutrino oscillation probabilities. A non-zero oscillation 
probability immediately leads to a violation of Bell's inequality
and to a teleportation fidelity value of greater than 2/3.

We first provide an introduction to the quantum mechanics of 
two flavour neutrino oscillations. Here we see that mode entanglement 
comes in a natural setting. We then discuss and compute different 
quantum correlations and relate them to the neutrino oscillation  
probabilities.  We finish with our conclusions.

%%%%%%%%%%%%%%%%%%%%%%%%%%%%%%%%%%%%%%%
\section{Quantum mechanics of two flavour neutrino oscillations}
%%%%%%%%%%%%%%%%%%%%%%%%%%%%%%%%%%%%%%%
It is well known that there are three flavour states of neutrinos,
$\nu_e, \nu_\mu$ and $\nu_\tau$ \cite{ALEPH:2005ab,Alexander:1991vi}. 
In the oscillation formalism, it is assumed that they mix via a 
$3 \times 3$ unitary matrix to form the three mass eigenstates 
$\nu_1, \nu_2$ and $\nu_3$. Neutrino oscillations occur only if the 
three corresponding masses, $m_1, m_2$ and $m_3$, are non-degenerate. 
Of the three mass-squared differences $\Delta_{kj} = m_k^2 - m_j^2$ 
(where $j,k = 1,2,3$ with $k > j$), only two are independent. 
Oscillation data tells us that $\Delta_{21} \approx 0.03 \times \Delta_{32}$,
hence $\Delta_{31} \approx \Delta_{32}$. One of the three mixing angles 
parametrizing the mixing matrix, $\theta_{13}$, is measured to be quite
small (about 0.14 radians) 
\cite{An:2012eh,Ahn:2012nd,Abe:2012tg,Dwyer:2013wqa}. 

In considering neutrino oscillations, in general, one should use the 
full three flavour oscillation formulae. A number of studies do this,
fitting all the available neutrino oscillation data to the three flavour
formulae \cite{Tortola:2012te,Fogli:2012ua,GonzalezGarcia:2012sz}. 
In the following three important experimentally relevant cases, 
the three flavour formula reduces to an effective two flavour formula:
\begin{enumerate}
\item
\underline{Long Baseline Reactor Experiments}: Reactors emit electron 
anti-neutrinos with energies of a few MeV. In long baseline reactor 
neutrino experiments, the baseline is expected to be greater than 50 km. 
For example, in KamLAND experiment \cite{Eguchi:2002dm,Araki:2004mb} 
the baseline is approximately 180 km. These experiments measure
the anti-neutrino survival probability $P(\bar{\nu}_e \to \bar{\nu}_e)$.
In the limit of neglecting $\theta_{13}$, this probability reduces to
the effective two flavour formula
[$1 - \sin^2 2 \theta_{12} \sin^2 (\Delta_{21}L/4E)$].
\item
\underline{Short Baseline Reactor Experiments}: These experiments have
baselines of about a km \cite{An:2012eh,Ahn:2012nd,Abe:2012tg,Dwyer:2013wqa}.
Given this short baseline, they are not capable to observing the 
oscillations induced by the smaller mass-square difference $\Delta_{21}$. 
Setting this quantity equal to zero in the expression for 
$P(\bar{\nu}_e \to \bar{\nu}_e)$, an effective two flavour formula  
[$1 - \sin^2 2 \theta_{13} \sin^2 (\Delta_{31}L/4E)$] is obtained
once again.
\item
\underline{Long Baseline Accelerator Experiments}: Accelerator 
neutrino beams consist of muon neutrinos (or anti-neutrinos) with 
energies ranging from hundreds of MeV \cite{Abe:2013fuq} to a few GeV 
\cite{Michael:2006rx}. They have baselines of hundreds of km. 
In the expression of the muon neutrino survival probability 
$P(\nu_\mu \to \nu_\mu)$ for these experiments, 
both the small parameters, $\Delta_{21}$ and $\theta_{13}$, can be
set to zero in the leading order. In this approximation, once again
an effective two flavour formula,  
[$1 - \sin^2 2 \theta_{23} \sin^2 (\Delta_{32}L/4E)$], is obtained. 
Then the problem reduces to that of two flavour 
mixing of $\nu_\mu$
and $\nu_\tau$ to form two mass eigenstates $\nu_2$ and $\nu_3$. 
The corresponding oscillations are described by one mixing angle
$\theta_{23}$ and one mass-squared difference $\Delta_{32}$.
\end{enumerate}

In the case of two flavour mixing, the relation between the flavour and the mass
eigenstates is described by a $2\times2$ rotation matrix, $U(\theta)$,
\be
\left(\begin{array}{c} \nu_{\alpha} \\ \nu_{\beta} \end{array}\right) = 
\left(\begin{array}{cc} \cos \theta & \sin \theta \\ - \sin \theta & \cos \theta \end{array}\right) 
\left(\begin{array}{c} \nu_j \\ \nu_k \end{array}\right)\,,
\label{ns}
\ee
where $\alpha,\,\beta =e,\, \mu$, $\tau$ and $j,\,k = 1,2,3$.
Therefore, each flavour state is given by a superposition of mass eigenstates,
\be
\ket{\nu_{\alpha}} = \sum_j U_{\alpha j} \ket{\nu_j}\,.
\ee

The time evolution of the mass eigenstates $\ket{\nu_{j}}$ is given by
\be
\ket{\nu_{j}(t)} = e^{-i E_{j} t} \ket{\nu_{j}}\,,
\ee
where $\ket{\nu_{j}}$ are the mass states at time $t=0$. Thus, we can write
\be
\ket{\nu_{\alpha} (t)} = \sum_j e^{-i E_j t} U_{\alpha j} \ket{\nu_j}\,.
\label{numass}
\ee

The evolving flavour neutrino state $\ket{\nu_{\alpha}}$ can also be projected on to the flavour basis
in the form 
\be
\ket{\nu_{\alpha}(t)} = \tilde{U}_{\alpha \alpha} (t)\ket{\nu_{\alpha}} + \tilde{U}_{\alpha \beta} (t)\ket{\nu_{\beta}}\,,
\label{t2}
\ee
where $\ket{\nu_{\alpha}}$ is the flavour state at time $t=0$ and 
$|\tilde{U}_{\alpha \alpha}(t)|^2+|\tilde{U}_{\alpha \beta}(t)|^2 = 1$.
We introduce occupation number states as \cite{neutri1,neutri2}
\be
\ket{\nu_{\alpha}}\equiv \ket{1}_{\alpha} \otimes \ket{0}_{\beta}\equiv \ket{10},\hspace{.16cm}
\ket{\nu_{\beta}}\equiv \ket{0}_{\alpha} \otimes \ket{1}_{\beta}\equiv \ket{01}\,.
\ee
Eq.~(\ref{t2}) can therefore be rewritten as
\be
\ket{\nu_{\alpha}(t)} = \tilde{U}_{\alpha \alpha} (t)\ket{1}_{\alpha} \otimes \ket{0}_{\beta} + \tilde{U}_{\alpha \beta} (t) \ket{0}_{\alpha} \otimes \ket{1}_{\beta}\,,
\label{flavmode}
\ee
where,
\bea 
\tilde{U}_{\alpha \alpha} (t) &=& \cos^2 \theta e^{-i E_j t} + \sin^2 \theta e^{-i E_k t}  ~, \nonumber\\
\tilde{U}_{\alpha \beta} (t) &=& \sin \theta \cos \theta (e^{-i E_k t} - e^{-i E_j t}) ~. 
\eea 
Now the state in Eq.~(\ref{flavmode}) has the form of a mode entangled single particle state \cite{enk,gerry,vedral,exp1,exp2}.
The corresponding density matrix is given by
\be
 \rho_{\alpha} (t) =
\left(\begin{array}{cccc} 
0 & 0 & 0 & 0 \\ 0 & |\tilde{U}_{\alpha \alpha} (t)|^2 & \tilde{U}_{\alpha \alpha} (t) \tilde{U^*}_{\alpha \beta} (t) & 0 \\ 
0 & \tilde{U}_{\alpha \beta} (t) \tilde{U^*}_{\alpha \alpha} (t)  & |\tilde{U}_{\alpha \beta} (t)|^2& 0 \\ 0 & 0 & 0 & 0
\end{array}\right)\,,
\label{dm}
\ee
where
\bea 
|\tilde{U}_{\alpha \alpha} (t)|^2  &=& c^4  + s^4 + 2 s^2 c^2 \cos \Big( \frac{\Delta_{}t}{2E} \Big) = P_{\rm sur}~, \label{psur} \\
|\tilde{U}_{\alpha \beta} (t)|^2 &=& 4 s^2 c^2 \sin^2 \Big( \frac{\Delta_{}t}{4E} \Big) = P_{\rm osc}~, \label{posc} \\
 \tilde{U}_{\alpha \alpha} (t) \tilde{U}^*_{\alpha \beta} (t) &=& s\, c \Big(s^2 - c^2 + c^2 e^{i \frac{\Delta_{}t}{2E}} - s^2 e^{-i \frac{\Delta_{}t}{2E}} \Big) ~, \\
%\nonumber\\
\tilde{U}_{\alpha \beta} (t) \tilde{U}^*_{\alpha \alpha}  (t) &=& s\, c\Big(s^2- c^2+ c^2 e^{-i \frac{\Delta_{}t}{2E}} - 
s^2 e^{i \frac{\Delta_{}t}{2E}} \Big)\,,
%\label{eq10}
\eea
with $c\equiv \cos \theta$ and $s \equiv \sin \theta$.
In the above equations, $\theta$ is a generic two flavour mixing 
angle and $\Delta$ is the correspoding mass-square difference.
Since the neutrino masses are very small (less than 1 eV), the neutrinos 
are assumed to be ultra relativistic. Hence the time of travel $t$ is
equal to the distance of travel $L$ and the difference in energies  
of the mass eigenstates $(E_k - E_j)$ can be set equal to $\Delta/2E$. 
The quantities in eqs.~(\ref{psur}) and~(\ref{posc}),
$|\tilde{U}_{\alpha \alpha}(t)|^2$ and $|\tilde{U}_{\alpha \beta}(t)|^2$, are
the two flavour survival and oscillation probabilities, respectively.
Note that $P_{\rm sur} < 1$, immediately implies $P_{\rm osc} > 0$.

%%%%%%%%%%%%%%%%%%%%%%%%%%%%%%%%%%%%%%%%%%%%%%%%%%%%%%%%
\section{Quantum correlations in two flavour neutrino oscillations}
\label{sec:def}
%%%%%%%%%%%%%%%%%%%%%%%%%%%%%%%%%%%%%%%%%%%%%%%%%%%%%%%%
In this section, we discuss and compute various quantum correlations inherent in the state
given in Eq.~(\ref{flavmode}). 
In all our subsequent calculations, the states considered are represented by $4 \times 4$ density matrices.

Bell's inequality is used to study the non-locality of a given system.
Its physical content is that a system that can be described by a local realistic 
theory will satisfy this inequality. Quantum mechanics 
provides many examples where this inequality gets violated \cite{aspect}. 
However, here we do not propose to derive a Bell's inequality from local realism.  
Instead we make use of a very interesting result obtained in \cite{hor1} 
which facilitates quantitative statements about Bell inequality violations
 just by making use of the parameters of the density  operator describing the system.

The density matrix $\rho$, in general, can be expanded in the form
\be
\rho=\frac{1}{4}[I\otimes I+(r.\sigma)\otimes I+I\otimes (s.\sigma)+\sum_{n,m=1}^3 T_{mn}(\sigma_m\otimes \sigma_n)]\,.
\ee
The elements of the correlation matrix $T$ are given by $T_{mn}=Tr\left[\rho(\sigma_m\otimes \sigma_n)\right]$. 
Let $u_i\ (i=1,2,3)$ be the eigenvalues of the matrix $T^{\dagger}T$. 
Then the Bell-CHSH inequality can be written as $M(\rho)\leq 1$, where $M(\rho)=\max(u_i+u_j)\ (i\neq j)$ \cite{hor1}. 
For the state (\ref{flavmode}), $M(\rho)$ is given by
\begin{eqnarray}
M(\rho) &=& 1 + \Big[3 + \cos4\theta + 2\cos\left(\frac{\Delta t}{2 E}\right) \sin^22\theta  \Big] \nonumber \\
&&\quad \times \sin^2 2\theta \sin^2\left(\frac{\Delta t}{4 E}\right), 
\nonumber \\
&=& 1 + 4 P_{\rm sur} P_{\rm osc}.
\end{eqnarray}
Thus we see that $M(\rho)$ is directly related to the neutrino 
oscillation probabilties and a measurement of $P_{\rm sur} < 1$ 
leads to a violation of Bell-CHSH inequality. We also note that
the maximal violation occurs when $P_{\rm sur} = 1/2 = P_{\rm osc}$.

Non-locality is the strongest aspect of quantum correlations. 
A weaker, though popular and extensively studied feature, is entanglement. 
For the case of entangled two-level systems it is synonymous with concurrence.
For a state with density matrix $\rho$, the concurrence is \cite{wootters}
\begin{eqnarray}
 C=\max(\lambda_1-\lambda_2-\lambda_3-\lambda_4,0),
\end{eqnarray}
where $\lambda_i$ are the square roots of the eigenvalues of $\rho \tilde{\rho}$  in decreasing order, where
  $\tilde{\rho}=(\sigma_y\otimes \sigma_y)\rho^*(\sigma_y\otimes \sigma_y)$ and is obtained by
applying the spin flip operation on $\rho$.  Here, concurrence can be shown to be 
\begin{eqnarray}
C &=& 2\sqrt{\sin^4\theta + \cos^4\theta+2\cos^2\theta \sin^2\theta\cos\left(\frac{\Delta t}{2 E}\right)} \nonumber \\
&& \times
\sin 2\theta \sin \left(\frac{\Delta t}{4 E}\right), 
\nonumber \\
&=& 2 \sqrt{P_{\rm sur} P_{\rm osc}} 
\end{eqnarray}
Obviously, entanglement is non-zero if the oscillation probability is
non-zero.

A still weaker measure of quantum correlations is quantum discord which points out that classicality and separability are not synonymous.
To obtain an analytical formula for quantum discord is a very difficult task as it
involves an optimization over local measurements, requiring numerical methods. To
overcome this difficulty, another measure of quantum correlation called geometric discord was introduced in \cite{dakic}
which quantifies the amount of non-classical correlation, of an
arbitrary quantum composite system, in terms of its minimal distance from the set of classical states. 
For $\rho$, geometric discord can be shown to be
\be
D_{G}(\rho)= \frac{1}{3}[\|\vec{y}\|^{2}+\|T\|^{2}-\lambda_{max}]\,,
\ee 
where $T$ is the correlation matrix defined above, $\vec y$ is the vector whose components are 
$y_{m}=\mathrm{Tr}(\rho(\sigma_{m}\otimes {{I}}_{}))$, and $\lambda_{max}$ is the maximum eigenvalue of the matrix $(\vec{y}\vec{y}^{\dagger}+ T T^{\dagger})$ \cite{dakic}.
It is not difficult to show that $D_{G}(\rho)$, here, is
\begin{eqnarray}
 D_{G}(\rho)&=& \frac{2}{3}  \Big[3 + \cos 4 \theta + 2 \cos \left(\frac{\Delta t}{2 E}\right) \sin^2 2 \theta \Big]
 \nonumber \\
 && \times \sin^2 2\theta \sin^2 \left(\frac{\Delta t}{4 E}\right)
 , \nonumber \\
&=&\frac{8}{3} P_{\rm sur} P_{\rm osc}.
\end{eqnarray}
 $D_{G}(\rho) \neq 0$ for $P_{\rm osc} \neq 0$, taking it away from
the classically allowed value of geometric discord \cite{ab12}.

Apart from the above foundational measures of various aspects of quantum correlations, a need was felt to have a measure that
defines the practical use of quantum correlations. This was supplied by teleportation. 
Since neutrinos interact only through weak interactions, 
the effect of decoherence is minimal, when compared to other particles 
such as electrons and photons that are widely used in quantum information 
processing. Hence it has the potential to impact practical quantum information processing.

The classical fidelity of teleportation in the absence of entanglement is $2/3$. Whenever the maximum teleportation fidelity, $F_{\max}>2/3$,  quantum teleportation is possible. 
$F_{\max}$, is easily computed in terms of the eigenvalues $\{u_i\}$ of $T^\dagger T$ mentioned above and is given by  $F_{\max}=\frac{1}{2}\left(1+\frac{1}{3}N(\rho)\right)$ where $N(\rho)=\left(\sqrt{u_1}+\sqrt{u_2}+\sqrt{u_3}\right)$ \cite{hor}. This expression allows for a useful interplay between teleportation fidelity and $M(\rho)$. This is so because $N(\rho)\geq M(\rho)$. Hence $M(\rho)>1$ automatically implies $F_{max}>2/3$. For the case of two flavour neutrino oscillations, $F_{\max}$ is given by 
\begin{eqnarray}
F_{\max} &=& \frac{2}{3} + \frac{1}{3}\sqrt{3+\cos 4\theta + 2\sin^2 2\theta \cos \left(\frac{\Delta t}{2 E}\right)} 
\nonumber \\
 && \quad \times
\sin 2 \theta \sin \left(\frac{\Delta t}{4 E}\right), \nonumber \\
&=&\frac{2}{3} \left( 1 + P_{\rm sur} P_{\rm osc} \right).
\end{eqnarray}
For non-zero $P_{\rm osc}$, $F_{\max} > 2/3$, where $2/3$ is the classical 
value of teleportation fidelity. Thus the usual relation between Bell's inequality 
violation and teleportation fidelity \cite{hor}, as seen in electronic and photonic 
systems, is obeyed here. This is in contrast to the unstable oscillating neutral mesons \cite{sbalok}.

 From the above analysis,  it is obvious that all the quantum correlations
are directly related to the neutrino oscillation probability. A measurement
of the neutrino survival probability which is less than unity directly
leads to the conclusion that all the quantum correlations take classically
forbidden values.

%%%%%%%%%%%%%%%%%%%%%%%%%%%%%%%%%%%%%%%
\section{Conclusions}
%%%%%%%%%%%%%%%%%%%%%%%%%%%%%%%%%%%%%%%

 In this work we have computed four facets of quantum correlations for the two 
flavour neutrino oscillations. We find that all these correlations are simple
functions of the product of neutrino survival and oscillation probabilities.
They acquire classically forbidden values when the oscillation probability 
is non-zero. In that case, the Bell's inequality is always violated and teleportation fidelity is always 
greater than 2/3. 
Since the three types of neutrino experiments
discussed in section~2, long and short baseline reactor and long baseline
accelerator, have all measured the neutrino survival probabilities to be
less than unity, we can conclude that they have also demonstrated the 
non-trivial quantum correlations in each case.

\end{document}